\documentstyle[aps,prl]{revtex}

\begin{document}
\twocolumn[
{\hsize\textwidth\columnwidth\hsize\csname@twocolumnfalse\endcsname
\draft
\title{Random interactions and spin-glass thermodynamic transition
in the hole-doped Haldane system Y$_{2-x}$Ca$_x$BaNiO$_5$}
\author{E. Janod, C. Payen, F.-X. Lannuzel, and K. Schoumacker}
\address{Institut des Mat\'{e}riaux Jean Rouxel, Universit\'{e} de
Nantes-CNRS,
BP 32229,
44322 Nantes Cedex 3, France}
\date{\today}
\maketitle

\begin{abstract}
Magnetization, DC and AC bulk susceptibility of the $S$=1 Haldane chain system doped with
electronic holes, Y$_{2-x}$Ca$_x$BaNiO$_5$ (0$\leq$x$\leq$0.20), have been
measured and analyzed. The most striking results are (i) a sub-Curie power law
behavior of the linear susceptibility, $\chi (T)$$\sim $ $T$$^{-\alpha }$, for temperature 
lower than the Haldane gap of the undoped compound (x=0) (ii) the existence of a 
spin-glass thermodynamic transition at $T$$_g$ = 2-3 K. These findings are consistent 
with (i) random couplings within the chains between the spin degrees of freedom induced by 
hole doping, (ii) the existence of ferromagnetic bonds that induce magnetic frustration
when interchain interactions come into play at low temperature.
\end{abstract}

}
]

Intensive research work is currently devoted to charge doping in Mott-Hubbard
and charge transfer insulators. In two- or three-dimensional (2D or 3D) transition metal
oxides, which are ordered antiferromagnets, hole doping yields properties ranging 
from colossal magnetoresistance in the manganates to high-T$_c$ superconductivity 
in the cuprates. In undoped one-dimensional (1D) compounds, enhanced quantum spin 
fluctuations may prevent the system from falling into a long-range ordered ground 
state (GS). Typical examples are the Heisenberg antiferromagnetic (AF) spin $S$=1 chain (Haldane chain) 
\cite{Haldane83} or the $S$=1/2 spin ladder \cite{Dagotto-Science}. Charge doping in this context occurs in 
a spin liquid GS. The nature of the GS emerging upon hole doping in these quantum disordered quasi-1D systems is clearly of 
interest. Due to quantum fluctuations, the GS could indeed be qualitatively different from the 
one appearing in 2D-3D systems \cite{Dagotto-Science}.

Among the existing quasi-1D systems, the Haldane chain compound Y$_2$BaNiO$_5$
is an attractive case. The quasi-1D character of its magnetic properties is related to its crystal
structure, made of chains of apex-sharing NiO$_6$ octahedra \cite{Buttrey90}. The two active 
e$_g$ orbitals associated with a strong ferromagnetic Hund's rule, result in a $S$=1 spin per 
nickel. The degeneracy of the energy levels (dx$^2$-y$^2$ and d3z$^2$-r$^2$) of the electrons 
composing the $S$=1 state is lifted due to the strong compression of the NiO$_6$ octahedra along 
the chain \cite{DiTusa94}. These spins couple together to form 1D $S$=1 AF chains with a superexchange coupling 
$J$ $\approx$ 280 K and a Haldane spin gap $\Delta$ $\approx$ 100 K \cite{Sakaguchi96}. In this compound, 
hole doping is achieved by an off-chain substitution, Ca$^{2+}$ for Y$^{3+}$ \cite{DiTusa94}.

On the experimental side, it was shown that the holes have O(2p$_z$)-Ni(3d$_{3z2-r2}$) character
\cite{DiTusa94,Maiti98,Lannuzel00} and their mobility induces a dramatic reduction of 
the resistivity along the $z$ chain axis. The charges are however essentially localized and a 
Variable-Range-Hopping-like temperature dependence of the resistivity is observed below 
room temperature \cite{DiTusa94}. Inelastic Neutron Scattering (INS) experiments reveal new magnetic
states below the Haldane gap that interact with each other \cite{DiTusa94,Xu-Science}. A spin
incommensurability (IC) appears for these subgap excitations with peaks symmetrically displaced 
about the q=$\pi$ point in the structure factor S(q). This spin IC is different from that seen in metallic
cuprates since the IC shift is found to be weakly doping-dependent. Xu et al. proposed that hole 
doping into the NiO chains induces a ferromagnetic (FM) interaction between the Ni spins on both 
sides of the hole (which is located on the apical O(2p$_z$) orbital) and that AF droplets develop around 
the hole \cite{Xu-Science}. Magnetic susceptibility data measured in the 2-20 K range by Kojima and 
co-workers \cite{Kojima95} give evidence for a spin-glass-like behavior below $T$$_f$ = 2.5 K and 
2.9 K for x=0.095 and 0.149, respectively. Muon spin rotation experiments indicate the presence of spin fluctuations 
persisting down to 50 mK, suggesting an unconventional ground state.

Assuming that the doped hole has a strong apical O(2p$_z$) character, several theoretical approaches 
\cite{Dagotto96,Batista98,Koshibae96} indicate the stabilization of a $S$=1/2 Zhang-Rice doublet (ZRD), 
due to a strong AF bond $J_K$ $\sim$ 1.4 eV between the residual $S$=1/2 spin on the oxygen in the chain 
and the $S$=1 spin of the nearest neighbor (NN) Ni$^{2+}$ ions. It leads to a strong FM interaction $J'$ $\sim$ 0.3 eV 
between the $S$=1/2 ZRD and the NN Ni$^{2+}$ $S$=1 ion, thus creating a polaron \cite{Batista98}. On the other 
hand, models assuming fully mobile holes with a purely Ni(3d) character predict the existence of FM 
polarons due to the double-exchange mechanism \cite{Ammon00}. Low energy magnetic excitations
exist within \cite{Batista98} and between polarons through the $S$=1 background \cite{Batista98,Ammon00} 
and thus explain the new magnetic states below the Haldane gap \cite{DiTusa94,Xu-Science}. The 
polarons were also shown to induce a magnetic frustration when interchain interactions come into play 
at low temperature \cite{Batista98}. Hereafter, all the spin degrees of freedom induced by a doped
hole, i.e. the spin of the hole, the FM bond and the AF droplets, will be referred to as 'polaron'.

In this context, we investigate in detail the bulk magnetic properties of Ca-doped Y$_2$BaNiO$_5$ through
magnetization, DC and AC susceptibilities measurements. In an intermediate temperature range 
8 K $\leq$ $T$ $\leq$ $\Delta$ = 100 K, a distribution of couplings between polarons generates a 
power-law behavior of the bulk susceptibility $\chi (T)$$\sim $ $T$$^{-\alpha }$ with $\alpha$
$\approx$ 0.6. A true thermodynamic transition into a spin-glass state appears at $T$$_g$ $\approx$ 2 - 3 K. 
The frustration associated with the spin glass state can be seen as the combined effect of ferromagnetic 
bonds created by hole doping and interchain couplings.

Polycrystalline samples of Y$_{2-x}$Ca$_x$BaNiO$_5$ with x=0, 0.05, 0.10, 0.15 and 0.20 were prepared
via solid-state reactions, as described elsewhere \cite{Alonso94}. X-ray diffraction characterization showed the
samples to be single-phased and very well crystallized. Linear variations of the crystal cell parameters
confirmed the existence of a homogeneous solid solution for x$\leq$0.20. Elemental microprobe analyses
were in excellent agreement with nominal compositions and showed very weak dispersion of the Ca
content for each doped sample. All AC and DC magnetization measurements were performed on a commercial
SQUID magnetometer (Quantum Design, U.S.A.) between 2 and 600 K.

The DC susceptibilities measured in a field of H=0.1 T, $\chi (T)$, of Y$_{2-x}$Ca$_x$BaNiO$_5$ (0$\leq$x$\leq$0.20) are shown 
in Fig. 1. The results obtained for the undoped sample (x=0) compare well with those in the published
literature \cite{Darriet93}. The $\chi (T)$ curve has a broad maximum at $T_{max}$ $\approx$ 380 K. The low-temperature
susceptibility (5 K$\leq$T$\leq$50 K) is well fitted by the sum of a T-independent term 
$\chi$$ _{0}$ $\approx$ $1.9 \times 10^{-4}$ cm$^{3}$ mol$^{-1}$, a Curie law $C/T$ with 
$C$ $\approx$ $6.3\times 10^{-3}$ cm$^{3}$ K mol$^{-1}$, and a thermally activated term with 
$\Delta$$\approx$100 K (see Fig. 1). The fitted value of $\chi_{0}$ agrees with the sum of the Van Vleck susceptibility
($\chi_{VV}$=(2.5$\pm 0.5$) $\times 10^{-4}$ cm$^{3}$ mol$^{-1}$ for a Ni$^{2+}$ ion with a $^{3}$A$_{1g}$
electronic ground term) and the diamagnetic contribution ($\chi_{dia}$ = $-1.2 \times 10^{-4}$ cm$^{3}$ mol$^{-1}$ 
\cite{Landolt-Borstein}). The Curie contribution corresponds to about 1.5 percent of free $S$=1/2 spins per formula unit. 
As seen in Fig. 1, the main effect of hole doping is an enhancement of the susceptibility $\chi (T)$. In 
the high temperature regime, $T$ $\geq$ $J$ = 280 K, the susceptibility is little
affected by doping. For $T < J$, a new regime appears where the enhancement of the susceptibility is stronger than 
for $T > J$. In particular, a non-Curie diverging behavior of $\chi (T)$ is observed for temperature smaller than the 
Haldane gap ($\Delta$=100 K) of the undoped compound. At low temperature, $T<5-10$ K, the divergence slows down 
and a saturation of the susceptibility is observed for the highest Ca contents. As already reported in Ref. \cite{Kojima95}, 
all specimens with $x>0.10$ 
show a spin-freezing above 2 K. This can be observed in the inset of Fig. 1 that displays zero-field-cooled (ZFC) and 
field-cooled (FC) susceptibilities for a static applied field of $H$= 0.005 T. A clear maximum appears in $\chi_{ZFC}$$(T)$
at $T$$_f$ = 2.5, 2.9 and 3.0 K for x=0.10, 0.15 and 0.20, respectively, with the magnetic irreversibility starting just 
below $T$$_f$. Based on the branching temperature between FC and ZFC data, $T$$_f$ decreases with increasing 
field for all samples, as observed in classical spin glass \cite{Binder86}. Another aspect of the spin-freezing was 
investigated through AC susceptibility measurements, as depicted in the inset of Fig. 1. The cusp in $\chi' (T)$ at 
$T$$_f$ is shifted towards lower temperature with decreasing frequency. The shift 
$\Delta$$T$$_f$/[$T$$_f$ $\Delta$(log $\omega$)] over three orders of magnitude in frequency is around 0.023, which 
is intermediate between the values observed for insulating and metallic canonical spin glasses, but far from those 
observed in superparamagnets \cite{Mydosh}.

In order to confirm that the magnetic irreversibility in $\chi (T)$ indicates a spin-glass thermodynamic transition, and not 
a simple dynamic freezing, we have analyzed the behavior of the nonlinear magnetization. In the critical regime 
near a spin glass transition, the magnetization $M$($H$,$T$) can be written as an expansion in odd powers of the
applied field $H$

\begin{eqnarray}
M(H,T) &=&\chi_{1} H - a_3 (\chi_{1} H)^{3} + a_{5} (\chi_{1} H)^{5}-...
\label{ecu2}
\end{eqnarray}
where $\chi_{1}$ is the linear susceptibility $M/H$ in the limit of $H \rightarrow 0$ and the coefficient $a_{2n+1}$ are 
functions of temperature \cite{Mydosh}. It is also useful to examine the nonlinear susceptibility,
$\chi_{nl}(H,T)$ ${= 1-M/(}$$\chi_{1}H)$ = - $a_3$($\chi_{1}H)$$^{2}$ + $a_5$($\chi_{1}H)$$^{4}$${  - ...}$ .
In the low field limit, the first term ($a_3$) in the expression of $\chi_{nl}(H,T)$ dominates. All nonlinear terms 
$a_{2n+1}$$(T)$ are expected to show power-law critical divergences upon approaching the critical spin-glass 
temperature $T_g$. Fig. 2 shows the $M(H,T)/H$ data of Y$_{1.8}$Ca$_{0.2}$BaNiO$_5$ (x=0.20) for several
temperatures above $T$$_f$ = 3.0 K. Data were obtained under field-cooled conditions. The static
field $H$ was switched on at $T$ $\approx$ 3 $T$$_f$ and kept constant during subsequent slow cooling down to the 
temperature of interest. We checked that data for x=0.1 and x=0.2 show the same qualitative behavior. The $M(H,T)$ 
data for x=0.20 have been fitted with the expression of Eq. 1 in the field range where the magnetic irreversibility is 
visible in the ZFC-FC set of measurements, i.e. $H$ $\leq$ 0.15 T, for $T$ $\geq$ 3.3 K (see Fig. 2). In this temperature range, 
the fitted values of the linear term, $\chi_{1}$, are the same as the $\chi_{FC}$ values obtained for H=0.005 T, to 
experimental uncertainties, and the $a_3$($T$) coefficient has a power-law divergence, 
$a_3$ $\sim$ $(T/T$$_f$ -1)$^{-\gamma}$ with $\gamma$ $\approx$ 1.6 and $T$$_f$ =3.0 K, in the range 
0.13$\leq$ $(T/T$$_f$ -1)$\leq$0.3. For $T$ $<$ 3.3 K, however, Eq. 1 seems to fail since a quadratic variation of $M(H,T)/H$ 
with $H$ is no longer observed (see Fig. 2). We looked for the reasons of this discrepancy by examining the nonlinear
susceptibility. Fig. 3 shows $\chi_{nl}(H,T)$, with $\chi_{1}$ extracted from the $H$=0.005 T FC measurements, as a function of
$H^2$. As one approaches $T_{f}$, the region where $\chi_{nl}$ is only linear in $H^2$ becomes much smaller due the
divergence of the higher order nonlinear terms, $a_{2n+1}$ $(n>1)$. This observation most probably explains why our data 
for $M/H$ are no longer consistent with a quadratic variation in $H$ for $T<3.3 K$.

The results depicted above clearly prove that the transitional aspects in doped Y$_2$BaNiO$_5$ are similar to those
existing in canonical spin-glass, supporting the existence of a true thermodynamic transition at $T_{f}$. Departures
from the standard phenomenology of spin-glasses are however observed below $T_{f}$. $\chi_{FC}(T)$ is 
temperature-dependent (Fig. 1), and the ratio $\chi'(0)/$$\chi'(T_{f}$) between the AC susceptibility at T=0 K and $T_{f}$ is close to 
0.95 against 0.5-0.6 in the usual cases \cite{Mydosh}. These behaviors could result from an unconventional (quantum) 
spin glass state. Persistent spin dynamics were indeed observed by muon spin relaxation \cite{Kojima95} much below 
$T_{f}$ in (Y,Ca)$_2$BaNiO$_5$, unlike in other standard spin glasses.

Let us turn now to the intermediate temperature range 8 K$<$T$<$$\Delta$=100 K. The off-chain Ca$^{2+}$ (hole) doping
has here different effects on $\chi (T)$ compared to in-chain non-magnetic doping (substitution Zn$^{2+}$ for Ni$^{2+}$)
\cite{Payen00}. The enhancement of $\chi (T)$ starts at much higher temperature and the divergence is slower in the case of
hole doping. This is in agreement with the distribution of the new magnetic states
inside the Haldane gap on a broad energy range for hole doping \cite{DiTusa94,Xu-Science} and restricted to the low energy region
for nonmagnetic doping \cite{DiTusa94,Monpean97}.

In a first phenomenological approach, we tried to describe the data using a modified Curie-Weiss law, 
$\chi (T)$=$\chi_{TI}$ + $C/(T$-$\theta$). In the range 10-50 K, the best fits yields $C$ $\approx$ 0.0316(3), 0.0837(1),
0.1247(4) and 0.1513(4) cm$^{3}$ K/Ba-mol, $\theta$ $\approx$ -1.7, -5.6, -7.4, -8.2 K for x=0.05, 0.10, 0.15, and 0.20, 
respectively. Our fitted Curie constants $C$ are significantly smaller than those found in Ref. \cite{Kojima95}. A model 
in which one Ca creates three free $S$=1/2 spins is therefore not obeyed in our data. The fitted values of $\chi_{TI}$ 
are almost doping-independent but are much higher than the $\chi_{0}$ term of the undoped compound 
($8.5 \times 10^{-4}$ against $1.9 \times 10^{-4}$ cm$^{3}$ mol$^{-1}$, see above). The validity of the modified Curie-Weiss 
model is therefore questionable.

Alternatively, it clearly appears in Fig. 1 that $\chi (T)$ can be described by a sub-Curie power-law over a significant temperature range
$T$ $<$ $\Delta$=100 K, regardless of the doping level. The temperature range where this dependence exists varies with the doping level,
from 8-25 K for x=0.05 to 20-75 K for x=0.20, as shown in the $\chi$ $T^{\alpha}$ vs $T$ plot in Fig. 4. We found a weak dependence
of $\alpha$ upon doping with $\alpha$ $\approx$ 0.595, 0.62, 0.64 and 0.655 for x=0.05, 0.10, 0.15 and 0.20, respectively. This 
behavior implies that new magnetic couplings are induced by doping, in full agreement with INS experiments \cite{DiTusa94,Xu-Science}.
Moreover, the broadening and the shift of the range where the power-law exists towards higher temperature with Ca content is a
strong indication that the average magnitude of the new magnetic couplings increases with doping. An explanation is that the
polarons interact with each other along the chain \cite{Xu-Science}. The magnetic couplings between polarons are exponentially small 
with distance because of the short-range order of the $S$=1 background (correlation length $\xi$ $\sim$ 7-8 \cite{Xu-Science}) which persists upon doping.
Their amplitude will therefore increase with doping since the average distance between neighboring polarons is greatly reduced at large 
Ca-content (down to 5 Ni-Ni distances for x=0.20). A broad random distribution of the couplings between polarons is moreover 
expected since no static charge order was detected in Y$_{2-x}$Ca$_x$BaNiO$_5$ \cite{Xu-Science}.

Interestingly a power-law divergence in $\chi (T)$ is also observed in quasi-1D random exchange Heisenberg systems such as the 
warwickites MgVOBO$_3$ ($S$=1 spins) \cite{Continentino96} and Mg$_{1-x}$Ti$_{1+x}$OBO$_3$ ($S$=1/2 spins)
\cite{Fernandes94}. In MgVOBO$_3$ ($\alpha$ $\approx$ 0.54), the Haldane gap is suppressed by a static bond disorder. 
The $\alpha$ exponent in Mg$_{1-x}$Ti$_{1+x}$OBO$_3$ ($\alpha$ $\approx$ 0.83) is independent of the concentration of Ti, 
i.e. of the disorder, in agreement with the properties of quasi-universality expected for a 1D random singlet ground state. 
Other bond disordered 1D $S$=1/2 systems such as some TCNQ salts \cite{Tippie81} however display a power-law behavior 
$\chi$ $\sim$ $T^{-\alpha}$ with a $\alpha$ exponent that decreases with the disorder, so these systems might be classified
as Griffiths paramagnets \cite{Hyman}. The physics of Y$_{2-x}$Ca$_x$BaNiO$_5$ is certainly different from the above examples 
since its behavior could be governed by polarons interacting through a $S$=1 medium, with a possible residual mobility of the polarons.
In this sense, Y$_{2-x}$Ca$_x$BaNiO$_5$ could correspond to a new type of 1D random exchange behavior.

Departures from the power-law are however observed at low temperature, as seen in Fig. 4. It is however difficult to
state whether it corresponds to pre-transitional effects or to the existence of a new energy scale. Interestingly, Ammon and
Imada suggested that fully mobile FM polarons may have a tendency to form pairs that share the magnetic disturbance induced by holes
\cite{Ammon00}. It would thus reduce the randomness and consequently the power-law behavior could disappear. Besides the 
power-law behavior, the polaron picture provides a microscopic explanation for magnetic frustration, since a polaron induces a $\pi$
shift of the staggered magnetization along the chain as depicted in Fig. 2 of Ref.\cite{Batista98}. Turning on the interchain
interactions therefore generates frustration, as suggested by Batista and co-workers \cite{Batista98}.

In summary, the main features of the static magnetic properties of the hole-doped Haldane chain Y$_{2-x}$Ca$_x$BaNiO$_5$
can be accounted for by the interplay of magnetic degrees of freedom induced by doping, interchain interactions and 
quantum fluctuations. In particular, we propose that the power-law dependence of $\chi (T)$ is a direct consequence of 
random couplings between polarons within the chain. We have also established that a true thermodynamic transition
into a spin-glass state occurs at low-temperature, pointing out the role played by interchain
interaction in determining the low-T behavior of doped Y$_2$BaNiO$_5$, as already suggested 
\cite{Batista98,Payen00,Batista98-Zn,Melin00}. 

We thank C.D. Batista,  A.A. Aligia and B. Ammon for helpful discussions.

\begin{figure}[tbp]
\caption{Magnetic susceptibility versus temperature, $\chi (T)$, for Y$_{2-x}$Ca$_x$BaNiO$_5$ 
(0$\leq$x$\leq$0.20), in the range 2-600 K ($H$=0.1 T). The solid line 
corresponds to a fit for x=0 (see text) and the dashed lines to a sub-Curie power law, $\chi$ $\sim$ $T^{-0.6}$.
Inset, lower part : DC zero-field-cooled and field-cooled parts in the range 2-4 K ($H$=0.005 T). Same symbols 
as in the main panel. Upper part: AC susceptibility $\chi' (T)$ for x=0.20, measured at $f$ = 1, 10,
99.9 and 999 Hz from the top to the bottom.}
\label{FIG. 1}
\end{figure}

\begin{figure}[tbp]
\caption{$M/H$ versus $H$ in Y$_{1.8}$Ca$_{0.2}$BaNiO$_5$ at several temperatures
above $T_{f}$=3.0 K as indicated. The solid lines are for the best fits with Eq. (1)}
\label{FIG. 2}
\end{figure}

\begin{figure}[tbp]
\caption{Non-linear susceptibility versus $H^{2}$ in Y$_{1.8}$Ca$_{0.2}$BaNiO$_5$ at several temperatures
above $T_{f}$=3.0 K as indicated.}
\label{FIG. 3}
\end{figure}

\begin{figure}[tbp]
\caption{$\chi T^{\alpha}$ versus $T$ in the intermediate temperature domain for
Y$_{2-x}$Ca$_x$BaNiO$_5$ (0.05$\leq$x$\leq$0.20). The horizontal lines are guides for the eyes.}
\label{FIG. 4}
\end{figure}

\end{document}